\documentclass[prd,aps]{revtex4}
\usepackage[dvips]{graphicx}
\begin{document}
\renewcommand{\baselinestretch}{1.5}
\newcommand{\Prd}{Phys. Rev D}
\newcommand{\Prl}{Phys. Rev. Lett.}
\newcommand{\Pl}{Phys. Lett.}
\newcommand{\Cqg}{Class. Quantum Grav.}
\newcommand{\Sch}{Schwarzschild$\;$}
\newcommand{\ochi}{\overline\chi}


\title{Effective Geometry}

\author {M. Novello and
S. E. Perez Bergliaffa}
\affiliation{Centro Brasileiro de Pesquisas F\'{\i}sicas, \\
Rua Dr.\ Xavier Sigaud 150, Urca 22290-180 Rio de Janeiro, RJ -- Brazil}

\begin{abstract}
\hfill{\small\bf Abstract}\hfill\smallskip
\par
We introduce the concept of effective geometry by studying several
systems in which it arises naturally. As an example of the power and conciseness
of the method, it is shown that
a flowing dielectric medium with a linear response to an external
electric field can be used to generate an analog geometry that has
many of the formal properties of a \Sch black hole for light rays,
in spite of birefringence. The surface gravity of this analog
black hole has a contribution that depends only on the dielectric
properties of the fluid (in addition to the usual term dependent
on the acceleration). This term may be give a hint to a new
mechanism to increase the temperature of Hawking radiation.
\end{abstract}
\maketitle
\newcommand{\beq}{\begin{equation}}
\newcommand{\eeq}{\end{equation}}
\newcommand{\vare}{\varepsilon}
\textbf{{\em Contribution to the Proceedings of the X$^{\rm th}$
Brazilian School of Gravitation and Cosmology, to be published by
AIP.} }
\section{Introduction}

In recent years, there has been a lot of interest in models that
mimic in the laboratory some features of gravitation
\cite{analog}. These models are built using systems that sometimes
look (deceivingly) simple, and are very different in nature: ordinary
nonviscous fluids, superfluids, flowing and non-flowing
dielectrics, non-linear electromagnetism in vacuum, and
Bose-Einstein condensates (see \cite{v1} for a complete list of
references). The underlying physics in all these cases is the
same: the behaviour of the fluctuations around a background
solution is governed by an ``effective metric''. More precisely,
the particles associated to the perturbations do not follow
geodesics of the background spacetime but of a Lorentzian geometry
described by the effective metric, which depends on the background
solution. This allows a rather complete analogy of some
kinematical aspects of general relativity \cite{visserprl}, but
not of its
dynamical features (see however \cite{v1,v2}).\\[.1cm]

Using this analogy, the geometrical tools of General
Relativity can be used to study some condensed matter systems
\cite{viss}. More important perhaps is the fact that the analogy
has permitted the simulation of several configurations of the
gravitational field, such as wormholes and closed space-like
curves for photons in nonlinear electrodynamics
\cite{biwh,ctc}, and warped spacetimes for
phonons \cite{mattuwe}. Particular attention has been paid to yet
another configuration, namely analog black holes, because these
would emit Hawking radiation exactly as gravitational black holes
do, and they are obviously much easier to generate in the
laboratory than their astrophysical counterpart. The fact that
analog black holes emit thermal radiation was shown first by Unruh
in the case of dumb black holes \cite{unruh}, and it is the
prospect of observing this radiation (thus testing the hypothesis
that the thermal emission is independent of the physics at
arbitrarily short wavelengths \cite{unruh}) that motivates the
quest for a realization of analog black holes in the laboratory.
Let us emphasize that the actual observation of the radiation is a
difficult task from the point of view of the experiment, if only
because of the extremely low temperatures involved. In the case of
a quasi one-dimensional flow of a Bose-Einstein condensate for
instance, the temperature of the radiation would be around 70 nK,
which is comparable but lower than
the temperature needed form the condensate \cite{matto}.\\[.1cm]

We shall begin by presenting in Sect.\ref{intro} the basics of the
idea of the effective geometry by giving a sketch of an example
given many years ago by W. Gordon \cite{gordon}. Then we shall
move on to the more interesting case of nonlinear
electromagnetism, where we introduce the mathematical tool
of {\em surface discontinuity}. In Sect. \ref{diel} we shall analyze another
example: photons in a flowing dielectric medium. We shall see
that, in analogy to the most general nonlinear electromagnetic
case, the photons experience bi-refringence {\em and}
bi-metricity. Then we demonstrate in Sect. \ref{abh} that it is
possible to build a static and spherically symmetric analog black
hole, generated by a {\em flowing} isotropic dielectric that
depends on an applied electric field. We give a specific example
in Sect.\ref{example}, in which the radius of the horizon and the
temperature depend on three parameters (the zeroth order
permittivity, the charge that generates the external field, and
the linear susceptibility) instead of depending only on the zeroth
order permittivity. As we shall show in Sect.\ref{new}, another
feature of this black hole is that there is a new term in the
surface gravity (and hence in the temperature of Hawking
radiation), in addition to the usual term proportional to the
acceleration of the fluid. This new term depends exclusively on
the dielectric properties of the fluid, and it might give an
opportunity to get Hawking radiation with temperature higher than
that reported up to date.

\section{The effective metric}
\label{intro}

Historically, the first example of the idea of effective metric was presented
by W. Gordon in 1923 \cite{gordon}. In modern language, the wave
equation for the propagation of light in a moving nondispersive
medium, with slowly varying refractive index $n$ and 4-velocity
$u^\mu$ is given by
$$
\left[ \partial_\alpha \partial^\alpha + (n^2-1) (u^\alpha
\partial_\alpha)^2 \right] F_{\mu\nu} = 0.
$$
Note that in this equation the components of $F_{\mu\nu}$ are not coupled.
Consequently, the propagation will be the same independently of the polarization. In
other words, there is no bi-refringence in moving media {\em with constant
refraction index} (we shall see later that this is not the case if
$n$ is a function of the coordinates).
Taking the geometrical optics limit, with the eikonal {\em Ansatz}, given by
$F_{\mu\nu} = {\cal F}_{\mu\nu}\;e^{i(\vec k . \vec x - \omega t)}$,
the Hamilton-Jacobi equation
for light rays can be written as  $ g^{\mu\nu}k_\mu k_\nu = 0 $
(see \cite{lp} for details), where
\beq g^{\mu\nu}= \eta^{\mu\nu}+
(n^2 - 1) u^\mu u^\nu \label{gordonm}
\eeq
is the effective metric
for this problem. It must be remarked that only photons in the
geometric optics approximation move on geodesics of $g^{\mu\nu}$:
the particles that compose the fluid couple instead to the
background Minkowskian metric (in fact, the dynamics of
the fluid is described by Euler's equation, and hence the
background spacetime seen by the fluid particles is Newtonian).

Let us study now in detail the example of nonlinear
electromagnetism. We start with the action
\beq
S = \int \sqrt{-\gamma}\;
L(F) \;d^{4}x ,\protect
\label{N1}
\end{equation}
where the invariant $F$ is given by
$F \equiv F^{\mu\nu}F_{\mu\nu}$ \footnote{We could have
considered $L = L(F,F^*)$ instead, where $F^* \equiv
F^{*}_{\mu\nu} F^{\mu\nu}$. This case is studied in
\cite{novelloplb}.}, and $L$ is an arbitrary function of
$F$. Notice that $\gamma$ is the determinant of the
background metric, which we take in the following to be that of
flat spacetime. However, the same techniques can be applied when the
background is curved (see for instance \cite{sns}). Varying this
action w.r.t. the potential $A_{\mu}$, related to the field by the
expression
$
F_{\mu\nu} =  A_{\mu,\nu} - A_{\nu,\mu},
$
we obtain the Euler-Lagrange equations of motion
(EOM):
\begin{equation}
(\sqrt{-\gamma}\;L_{F} F^{\mu\nu})_{;\nu} = 0,
 \protect\label{N2}
\end{equation}
where $L_{F}$ is the functional derivative $ L_{F} \equiv \frac{\delta L}{\delta F}.$
In the particular case of a linear dependence of the Lagrangian with
the invariant $F$ we recover Maxwell's EOM.

As mentioned in the Introduction, we would like to study
the behaviour of perturbations of these EOM around a fixed background solution. In particular,
we shall be interested in the causal structure inherent to the EOM (\ref{N2}). This structure
is described by the characteristics of the EOM \cite{ch}. Instead of
using the infinite-momentum limit of the
eikonal approximation \cite{mattfest}, we shall use
a more elegant method set out by Hadamard \cite{hadamard}. In this method, the propagation
of low-energy photons is studied by following the evolution of the wave front
({\em i.e.} the characteristic surface), through
which the electromagnetic
field is continuous but its first derivative is not. To be specific,
let $\Sigma$ be the surface of discontinuity defined by the equation
$$\Sigma(x^{\mu}) = {\rm constant}.$$
The discontinuity of a function $J$ through $\Sigma$ will be represented by
$[J]_\Sigma$, and its definition is
$$
[J]_\Sigma \equiv \lim_{\delta\rightarrow 0^+} \left( \left. J\right|_{\Sigma +\delta}
- \left. J \right|_{\Sigma - \delta}\right) .
$$
The discontinuities of the field and its first
derivative are given by
\begin{equation}
[F_{\mu\nu}]_{\Sigma} = 0,   \;\;\;\;\;\;\;\;\;\;\;\;\;\;\;\;\;\;\;\;\;\;
[F_{\mu\nu,\lambda}]_{\Sigma} = f_{\mu\nu}  k_{\lambda},
\protect\label{N8}
\end{equation}
where the vector $k_{\lambda}$ is nothing but the normal to the surface
$\Sigma$, that is, $k_{\lambda} = \Sigma_{,\lambda}.$

To set the stage for the nonlinear case, let us first discuss the causal properties of
Maxwell's electrodynamics,
for which $ L_{F} = $const.
The EOM then reduces to $F^{\mu\nu}_{\;\;,\nu} = 0$, and taking its
discontinuity we get
\begin{equation}
f^{\mu\nu}  k_{\nu} = 0.
\protect\label{N10}
\end{equation}
The other Maxwell equation is given by ${F^{*}_{\mu\nu}}^{,\nu} = 0$
or equivalently,
\begin{equation}
F_{\mu\nu,\lambda} + F_{\nu\lambda,\mu} + F_{\lambda\mu,\nu} = 0.
\protect\label{N12}
\end{equation}
The discontinuity of this equation yields
\begin{equation}
f_{\mu\nu}  k_{\lambda} + f_{\nu\lambda}  k_{\mu} + f_{\lambda\mu}  k_{\nu} = 0.
\protect\label{N13}
\end{equation}
Multiplying this equation by $k^{\lambda}$ gives
\begin{equation}
f_{\mu\nu}  k^2 + f_{\nu\lambda}  k^{\lambda} k_{\mu} + f_{\lambda\mu}
k^{\lambda} k_{\nu} = 0,
\protect\label{N14}
\end{equation}
where $k^2 \equiv k_{\mu} k_{\nu} \gamma^{\mu\nu}. $
Using the orthogonality condition from Eqn.(\ref{N10}) it follows that
\begin{equation}
f^{\mu\nu}  k^2  = 0.
\protect\label{N15}
\end{equation}
Since the tensor associated to the discontinuity cannot vanish (we are assuming that
there is a true discontinuity!) we conclude that the surface of
discontinuity is null w.r.t. the metric $\gamma^{\mu\nu}$. That is,
\begin{equation}
k_{\mu} k_{\nu} \gamma^{\mu\nu} = 0.
\protect\label{N16}
\end{equation}
(compare with Eqn.(\ref{gordonm})).
It follows that
$ k_{\lambda,\mu} k^{\lambda} = 0$,
and since the vector of discontinuity is a gradient,
\begin{equation}
k_{\mu,\lambda} k^{\lambda} = 0.
\protect\label{N17}
\end{equation}
This shows that the propagation of discontinuities of the electromagnetic
field, in the case of Maxwell's equations (which are linear), is
along the null geodesics of the Minkowski background metric.\\[.1cm]

Let us apply the same technique to the case of a
nonlinear Lagrangian for the electromagnetic field, given by $L(F)$. Taking the
discontinuity of the EOM (\ref{N2}), we get
\begin{equation}
L_{F} f^{\mu\nu} k_{\nu} + 2 a\;L_{FF}\;  F^{\mu\nu} k_{\nu} = 0,
\protect\label{N18}
\end{equation}
where we defined the quantity $a$ by $ F^{\alpha\beta}f_{\alpha\beta} \equiv a. $
Note that contrary to the linear case in which the discontinuity
tensor $f_{\mu\nu}$ is orthogonal to the propagation vector $k^{\mu}$, here
there is a complicated relation between the vector  $f^{\mu\nu} k_{\nu}$ and
quantities dependent on the background field. This is the origin of a more
involved expression for the evolution of the discontinuity vector, as we
shall see next.
Multiplying equation (\ref{N14}) by $F^{\mu\nu}$ we obtain
\begin{equation}
a  \;k^2 + F^{\mu\nu} f_{\nu\lambda}  k^{\lambda} k_{\mu} +
F^{\mu\nu} f_{\lambda\mu}  k^{\lambda} k_{\nu} = 0.
\protect\label{N141}
\end{equation}
Now we substitute in this equation the term $  f^{\mu\nu} k_{\nu} $ from Eqn.(\ref{N18}),
and we arrive at the expression
\begin{equation}
a k^2 - 2 \frac{L_{FF}}{L_{F}} a ( F^{\mu\lambda} k_{\mu} k_{\lambda} -
 F^{\lambda\mu} k_{\mu} k_{\lambda} ),
\protect\label{N19}
\end{equation}
which can be written as $g^{\mu\nu}k_\mu k_\nu = 0$, where
\begin{equation}
g^{\mu\nu}= \gamma^{\mu\nu} - 4 \frac{L_{FF}}{L_{F}} F^{\mu\nu}.
\protect\label{N20}
\end{equation}
We then conclude that\\[.1cm]
\begin{center}%
\fbox{\fbox{\parbox{12cm}{
The low-energy photons of a {\em nonlinear} theory of electrodynamics
with $L=L(F)$ do not propagate on the null cones of the background metric but
on the null
cones of an {\em effective} metric, generated by the self-interaction of the
electromagnetic field.
}}}
\end{center}
\vspace{.2cm}

This statement is always true in case of Lagrangians depending only of the invariant $F$.
For Lagrangians that depend also of $F^*$, there may be
some special cases in which the propagation
coincides with that in Minkowski \cite{novelloplb}.
Another feature of the more general case $L=L(F,F^*)$ is that bi-refringence is present.
That is, each of the two polarization states of the photon has its own dispersion relation.
In some special
cases, there is also bi-metricity (one effective metric for each polarization
state). Some more
special cases (such as Born-Infeld electrodynamics) even
exhibit only a single metric \cite{n1,mattfest}. Several of these features are present in our next
example.

\section{Effective metric(s) in the presence of a dielectric}
\label{diel}

Let us now move
to another interesting case where the effective geometry is useful to study the
causal properties of low-energy photons. We shall analyze
the propagation of such photons in a
nonlinear medium (see Ref.\cite{cqg} for details and notation).
Let us define first the antisymmetric tensors $F_{\mu\nu}$ and
$P_{\mu\nu}$, which are convenient to represent the electromagnetic field
when material media are present.
These tensors can be expressed in terms of the strengths ($E$, $H$) and
the excitations ($D$, $B$) of the electric and magnetic fields as
\begin{eqnarray*}
F_{\mu\nu} &=& v_{\mu}E_{\nu} - v_{\nu}E_{\mu}
- \eta_{\mu\nu}{}^{\alpha\beta}v_{\alpha}B_{\beta},
\label{1}
\\
P_{\mu\nu} &=& v_{\mu}D_{\nu} - v_{\nu}D_{\mu}
- \eta_{\mu\nu}{}^{\alpha\beta}v_{\alpha}H_{\beta} .
\label{2}
\end{eqnarray*}
where $v_\mu$ represents the 4-velocity of an
arbitrary observer (which we will take later as comoving
with the fluid). The Levi-Civita tensor
introduced above is defined in such way that $\eta^{0123} = +1$ in Cartesian coordinates.
Since the electric and magnetic fields are spacelike vectors,
we shall use the notation
\begin{math}
E^{\alpha}E_{\alpha}\equiv -E^{2},
H^{\alpha}H_{\alpha}\equiv -H^{2}.
\end{math}
We will consider here media with properties determined only by the tensors
$\epsilon_{\alpha\beta}$ and $\mu_{\alpha\beta}$ ({\em i.e.} media with
null magneto-electric tensor), which relate the electromagnetic excitations
to the field strengths by the constitutive laws,
\beq
D_{\alpha} =  \epsilon_{\alpha}{}^{\beta}(E,H)E_{\beta},\;\;\;\;\;\;\;\;
B_{\alpha} = \mu_{\alpha}{}^{\beta}(E,H)H_{\beta}.
\label{consteq}
\eeq
In order to get the effective metric, we shall use Hadamard's method \cite{hadamard} as
in the previous section.
By taking the discontinuity of the field equations
$^*F^{\mu\nu}{}_{,\nu}=0$ and $P^{\mu\nu}{}_{,\nu}=0$, and assuming that
\beq
\epsilon^{\mu\beta} = \epsilon(E)(\gamma^{\mu\beta}-v^{\mu}v^{\beta}),
\label{epsilon}
\end{equation}
and
\beq
\mu^{\mu\beta} = \mu_0(\gamma^{\mu\beta}-v^{\mu}v^{\beta}),
\label{mu0}
\end{equation}
with $\mu_0 = $ const., we get the following equations:
\beq
\epsilon (k.e) - \frac{\epsilon'}{E} (E.e)(k.E) = 0,
\label{ke}
\eeq
\beq
\mu_0 (k.h)=0,
\eeq
\beq
\epsilon (k.v) e^\mu - \frac{\epsilon'}{E} E^\alpha e_\alpha (k.v) E^\mu +
\eta^{\mu\nu\alpha\beta} k_\nu v_\alpha h_\beta = 0,
\label{eom3}
\eeq
\beq
\mu_0 (k.v) h^\mu - \eta^{\mu\nu\alpha\beta} k_\nu v_\alpha e_\beta = 0,
\label{eom4}
\eeq
where $k^\mu$ is the wave propagation vector, $\epsilon '$ is the derivative of $\epsilon$ w.r.t.
$E$, and
$$
[E_{\mu ,\lambda}]_\Sigma = e_\mu\;k_\lambda ,\;\;\;\;\;\;\;\; [H_{\mu ,\lambda}]_\Sigma
= h_\mu\;k_\lambda .
$$
Note in particular that Eqn.(\ref{ke}) shows that the vectors $k^\mu$ and $e^\mu$ are not
always orthogonal, as would be the case if $\epsilon '$ was zero.
Substituting Eqn.(\ref{eom4}) in (\ref{eom3}), we get
\beq
Z^{\mu\beta}e_{\beta} = 0,
\label{53}
\eeq
where the matrix $Z$ is given by
\beq
Z^{\mu\beta}  =  \left[ k^2 + (k.v)^2 (\mu_0 \epsilon -1) \right]
\gamma^{\mu\beta}
 -\mu_0 \frac{\epsilon'}{E} (k.v)^2 E^\mu E^\beta
 + (v.k) (v^\mu k^\beta + k^\mu v^\beta )
  - \left[ \epsilon \mu_0 (k.v) +k^2 \right]\;
v^\mu v^\beta - k^\mu k^\beta  .
\label{54}
\eeq
Non-trivial solutions of Eqn.(\ref{53}) can be found only for
cases in which $\det\left| Z^{\mu\beta} \right| = 0$ (
this condition is a generalization of the well-known Fresnel equation
\cite{Landau}).\\[.1cm]

Eqn.(\ref{53}) can be solved by expanding
$e_{\nu}$ as a linear combination of the four linearly independent
vectors $v_{\nu}$, $E_{\nu}$, $k_{\nu}$ and
$\eta_{\alpha\beta\mu\nu}v^\alpha E^\beta k^\mu$ \footnote{The particular
instance in which the vectors used as a basis in
Eqn.(\ref{e-mu}) are not linearly independent is discussed in \cite{cqg}.}.
That is,
\begin{equation}
e_{\nu}=\alpha E_{\nu}
+\beta\eta_{\alpha\lambda\mu\nu}v^\alpha E^\lambda k^\mu
+\gamma k_{\nu}+\delta v_{\nu}.
\label{e-mu}
\end{equation}
Notice that taking the discontinuity of $E^\mu_{\;\;,\lambda}$ we can show that
$(e.v) = 0$. This restriction imposes a relation between the coefficients of
Eqn.(\ref{e-mu}):
$$
\delta = -\gamma (k.v)
$$
With the expression given in Eqn.(\ref{e-mu}), Eqn.(\ref{53}) reads
\begin{eqnarray}
\alpha\left[
k^2 -\left(1 - \mu_0\;(\epsilon \; E)'
\right)(k.v)^2 \right] -\gamma\left[ \mu_0 (k.v)^2\frac{1}{E}
\;\epsilon '
E^{\alpha}k_{\alpha} \right] &=& 0 ,\nonumber
\\
\alpha E^{\mu}k_{\mu}+\gamma(1-\mu_0\epsilon)(k.v)^2+\delta(k.v) = 0, & & \nonumber
\\
\alpha(k.v)E^{\mu}k_{\mu}+\gamma(k.v)k^2+
\delta\left[k^2+\mu_0\epsilon\;(k.v)^2 \right]= 0 ,& & \nonumber
\\
\beta
\left[k^2-(1-\mu_0\epsilon)(k.v)^2\right] = 0. & &  \nonumber
\end{eqnarray}
The solution of this system results
in the following dispersion relations:
\begin{eqnarray}
k_-^2 &=& (k.v)^2\left[1 - \mu_0(\epsilon\;  E)'
\right]
+ \frac{1}{\epsilon E}\;\epsilon'\;
E^{\alpha}E^{\beta}k_{\alpha}k_{\beta},
\label{g-5}\\
k_+^2 &=& [1-\mu_0\epsilon(E)](k.v)^2.
\label{g-6}
\end{eqnarray}
They correspond to the propagation modes
\begin{eqnarray}
e^-_\nu&=&\rho^- \left\{\mu_0\;\epsilon(k.v)^2 E_\nu
+E^\alpha k_\alpha[k_\nu-(k.v)v_\nu]\right\} ,
\label{e-}\\
e^+_\nu&=&\rho^+ \;\eta_{\alpha\lambda\mu\nu}v^\alpha E^\lambda k^\mu,
\label{e+}
\end{eqnarray}
where $\rho^-$ and $\rho^+$ are arbitrary constants.
The labels ``$+$'' and ``$-$'' refer to the ordinary and extraordinary rays,
respectively.
Eqns.({\ref{g-5}) and ({\ref{g-6}) govern the propagation of photons in the medium characterized
by $\mu = \mu_0 = $const., and $\epsilon =
\epsilon (E)$. They can be
rewritten as $g_{\pm}^{\mu\nu}k_{\mu}k_{\nu}=0$,
where we have defined the effective geometries
\begin{eqnarray}
g_{(-)}^{\mu\nu} &=& \gamma^{\mu\nu} -
\left[1 - \mu_0\; (\epsilon\; E)'\right]v^{\mu}v^{\nu}-\frac{1}{\epsilon E}
\;\epsilon '\;E^{\mu}E^{\nu},
\label{h-2}\\
g_{(+)}^{\mu\nu} &=& \gamma^{\mu\nu}-[1-\mu_0\;\epsilon]v^\mu v^\nu.
\label{glmetric}
\end{eqnarray}
The metric given by Eqn.(\ref{h-2}) was derived first in \cite{ns1}, while
the second metric very much resembles the metric obtained by Gordon \cite{gordon}
(see Eqn.(\ref{gordonm}). The
difference is that in the case under consideration,
$\epsilon$ is a function of the modulus of the external electric field,
while Gordon worked with a constant
permeability.

We see then that in this example each polarization state has its own dispersion relation
(Eqns.(\ref{g-5}) and (\ref{g-6})), so there is bi-refringence. There is also bi-metricity, because
each type of photon moves according a different metric
(see Eqns.(\ref{h-2}) and (\ref{glmetric})).

\section{The Analog Black Hole}
\label{abh}

We shall show in this section that the system described
by the effective
metrics given by Eqns.(\ref{h-2})-(\ref{glmetric}) can be used to produce
an analog black hole.
It will be convenient to rewrite at this point the inverse of the
effective metric given by Eqn.(\ref{h-2}) using a
different notation:
\beq
g_{\mu\nu}^{(-)} = \gamma_{\mu\nu} - \frac{v_\mu v_\nu}{c^2} (1 - f) +
\frac{\xi}{1+\xi} \; l_\mu l_\nu ,
\label{nsmetric}
\eeq
where we have defined the quantities
$$f \equiv \frac{1}{c^2\mu_0\epsilon (1+\xi)},\;\;\;\;\;\;\xi \equiv
 \frac{\epsilon ' E}{\epsilon},
\;\;\;\;\;\;l_\mu \equiv \frac{E_\mu}{E}.
$$
Note that
$\epsilon = \epsilon (E)$. We have introduced here the velocity of light $c$, which was
set to 1 before.
Taking a Minkowskian background in spherical coordinates, and
\beq
v_\mu = (v_0 , v_1 , 0 , 0 ), \;\;\;\;\;\;\; E_\mu =
(E_0,E_1, 0 , 0 ),
\label{vE}
\eeq
we get for the effective metric described by Eqn.(\ref{nsmetric}),
\beq
g_{00}^{(-)} = 1 - \frac{v_0^2}{c^2}\; (1 - f) +
\frac{\xi}{1+\xi} \; l_0^2 ,
\eeq
\beq g_{11}^{(-)} = -1 -\frac{v_1^2}{c^2}\; (1 - f ) +
\frac{\xi}{1+\xi} \; l_1^2 ,
\eeq
\beq
g_{01}^{(-)} = -\frac{v_0 v_1}{c^2} \;(1 - f) +
\frac{\xi}{1+\xi}\; l_0\; l_1 ,
\eeq
and $g_{22}^{(-)}$ and $g_{33}^{(-)}$ as in
Minkowski spacetime.
The vectors $v_\mu$ and $l_\mu$ satisfy the constraints
\beq
v_0^2 - v_1^2 = c^2,
\label{c4}
\eeq
\beq
l_0^2 - l_1^2 = -1,
\label{c3}
\eeq
\beq
v_0 l_0 - v_1 l_1 = 0.
\label{c2}
\eeq
This system of equations can be solved in terms of  $v_1$, and the result is
\beq
v_0^2 = c^2 + v_1^2 ,
\label{v0}
\eeq
\beq
l_0^2 = \frac{v_1^2}{c^2} ,\;\;\;\;\;\;\;\;\;\;\;\; l_1^2 = \frac{c^2+v_1^2}{c^2}.
\label{l0}
\eeq

Now we can rewrite the metric in terms of $\beta\equiv v_1/c$, a definition
which
coincides with the usual one for small values of $v_1$.
The explicit expression of the metric coefficients is:
\beq
g_{00}^{(-)} = \frac{1 - \beta^2(c^2\mu_0 \epsilon -1)}{c^2\mu_0 (\epsilon + \epsilon
' E)}
\label{g00},
\eeq
\beq
g_{01}^{(-)} = \beta  \sqrt{1+\beta^2}\; \frac{1-c^2\mu_0\epsilon}{c^2\mu_0
(\epsilon +\epsilon ' E)} ,
\label{g01}
\eeq
\beq
g_{11}^{(-)} = \frac{\beta^2 -c^2\mu_0\epsilon (1+\beta^2)}
{c^2\mu_0 (\epsilon + \epsilon ' E)} .
\label{g11}
\eeq
From Eqn.(\ref{g00}) it is easily seen that, depending on the function
$\epsilon(E)$, this metric has a horizon at $r=r_h$, given
by the condition $g_{00}(r_h) = 0$ or equivalently,
\beq
\left. \left( c^2\mu_0 \epsilon - \frac{1}{\beta^2}\right) \;\right|_{r_h} = 1 .
\label{rh}
\eeq
The metric given above resembles the form of
Schwarzschild's solution in Painlev\'e-Gullstrand coordinates
\cite{pain,gull}:
\beq
ds^2 = \left(1 - \frac{2GM}{r}\right) dt^2 \pm 2\sqrt{\frac{2GM}{r}}\;dr\;dt - dr^2 - r^2 d\Omega^2 .
\label{pain}
\eeq
With the coordinate transformation
\beq
dt_{P} = dt_S \mp \frac{\sqrt{2GM/r}}{1-\frac{2GM}{r}} \;dr ,
\label{trafo}
\eeq
the line element given in Eqn.(\ref{pain}) can be written in Schwarzschild's coordinates. The
``$+$'' sign covers the future horizon and the black hole singularity.

The effective metric given by Eqns.(\ref{g00})-(\ref{g11}) looks like the metric in
Eqn.(\ref{pain}) \footnote{Note that a conformal factor to make $g_{11}=-1$ in Eq.(\ref{g11})
is needed. Consequently, the two metrics are actually conformally equivalent.}.
In fact, it can be
written in
Schwarzschild's coordinates, with the coordinate change
\beq
dt_{PG}  = dt_S - \frac{g_{01}(r)}{g_{00}(r)} dr .
\label{tr}
\eeq
Using this transformation with the metric coefficients given in Eqns.(\ref{g00}) and (\ref{g01}),
we get the expression of $g_{11}^{(-)}$ in \Sch coordinates:
\beq
g_{11}^{(-)} = -\frac{\epsilon(E)}{(1-\beta^2[c^2\mu_0\epsilon(E) -1])(\epsilon(E) + \epsilon(E) 'E)}.
\eeq
Note that $g_{01}^{(-)}$ is
zero in the new coordinate system, while
$g_{00}^{(-)}$ is still given by Eqn.(\ref{g00}). Consequently,
the position of the horizon does not change, and is still given by Eqn.(\ref{rh}).
\\[.1cm]

Working in Painlev\'e-Gullstrand coordinates, we have shown that
the metric for the ``$-$'' polarization describes a Schwarzschild black hole
if Eqn.(\ref{rh}) has a solution. Afterwards we have
rewritten the ``$-$'' metric in more familiar coordinates.
By means of similar calculations, it can be shown that
photons with the other polarization ``see''
the metric (in \Sch coordinates) given by
\beq
g_{00}^{(+)} = \frac{1+\beta^2(1-c^2\mu_0\epsilon(E))}{c^2\mu_0\epsilon(E)},
\label{g00m}
\eeq
\beq
g_{11}^{(+)} = -\frac{1}{1+ \beta^2 (1-c^2\mu_0 \epsilon(E))}.
\eeq
\vspace{.1cm}

It is important to stress then
that {\em the horizon is located at $r_h$ given by
Eqn.(\ref{rh}) for photons with any polarization}.
Moreover, the motion of the photons in both geometries
will be qualitatively the same, as we shall show
below.

\section{An example}
\label{example}

We have not specified up to now the functions $\epsilon(E)$ and
$E(r)$ that determine the dependence of the coefficients of the
effective metrics with the coordinate $r$. From now on we assume a
linear $\epsilon(E)$, a type of behaviour which is
exhibited for instance by electrorheological fluids \cite{erf}. Specifically, we take
\beq
\epsilon(E) = \epsilon_0 (\overline\chi +
\chi^{(2)} E(r)) ,
\label{eps}
\eeq
with
$\overline\chi =
1+\chi^{(1)}$. The nontrivial Maxwell's equation then reads
\beq \left(
\sqrt{-\gamma}\;\epsilon (r) F^{01}\right)_{,1} = 0.
\label{max}
\eeq
Taking into account that $ (F^{01})^2 =
\frac{E^2}{c^2},$ we get
\beq
F^{01} = \frac{-\overline\chi \pm \sqrt{\ochi^2 + 4\chi^{(2)}
Q/\epsilon_0 r^2}}{2c\chi^{(2)}} .
\eeq
as a solution of Eqn.(\ref{max}) for
a point source in a flat background in spherical coordinates.
Let us consider a
particular combination of parameters: $\chi^{(2)} >0$, $Q > 0$ and
the ``$+$'' sign in front of the square root in $F^{01}$, in such
a way that $E>0$ for all $r$. To get more manageable expressions
for the metric, it is convenient to define the function $\sigma
(r)$: \beq E(r) \equiv \frac{\ochi}{2\chi^{(2)}}\; \sigma (r) \eeq
where \beq \sigma (r) = -1 + \frac{1}{r} \sqrt{r^2 + q}
\label{sigma}
\eeq
and
\beq
q = \frac{4\chi^{(2)} Q}{\epsilon_0\ochi ^2 }.
\label{q}
\eeq
In terms of $\sigma$, the metrics take the form
\beq
ds_{(-)}^2  =  \frac{2-\beta^2\;[\;\ochi\; (\sigma (r)+2) -2]}{2\;\ochi\;
 (1+\sigma(r))}\; d\tau ^2 -
 \frac{2+\sigma (r)}
{[2-\beta^2\;(\ochi\; (\sigma (r) +2)-2)]\;(1+\sigma (r))}
\;dr^2 - r^2 d\Omega^2,
\label{nsm}
\eeq
\beq
ds_{(+)}^2  =
 \frac{2-\beta^2\;[ \;\ochi\;(\sigma (r) +2)-2]}{\ochi\; (2+\sigma (r))}\; d\tau^2
 -
  \frac{2}{2+\beta^2\;[2-\ochi\; (\sigma (r)+2)]}\; dr^2 - r^2 d\Omega^2.
\label{gm}
\eeq
Notice that the $(t,r)$ sectors of these metrics
are related by the following expression:
\beq
ds^2_{(+)} = \Phi (r)\; ds^2_{(-)}
\label{confinv}
\eeq
where the conformal factor $\Phi$ is given by:
$$
\Phi = 2\;\frac{1+\sigma (r)}{2+\sigma (r)}
$$

We shall study next some features of the effective black hole metrics.
It is important to remark that up to this point, the velocity of the fluid
$v_1$ is completely
arbitrary; it can even be a function of the coordinate $r$. We shall assume in the following
that $v_1$ is a constant. This assumption, which will be lifted in
Sect.\ref{new}, may seem rather restrictive but it helps to
display the main features of the effective metrics in an easy way.

To study the motion of the photons in these geometries, we can use the technique of
the effective potential.
Standard manipulations (see for instance \cite{wald}) show that in the case of a
static and spherically symmetric metric, the effective potential is given by
\beq
V(r) = \varepsilon^2 \left( 1 + \frac{1}{g_{00}(r)\;g_{11}(r)}\right) - \frac{L^2}{r^2
g_{11}(r)}
\label{effpot}
\eeq
where $\varepsilon$ is the energy and $L$ the angular momentum of the photon.

In terms of $\sigma (r)$, and of the impact parameter $b^2 = L^2/\varepsilon^2$, the
"small" effective potential $v(r)\equiv V(r)/\varepsilon^2$ for the metric Eqn.(\ref{nsm})
 in \Sch coordinates can be written as follows:
\beq
v^{(-)}(r)  =  1-\frac{2(1+\sigma (r))^2}{2+\sigma (r)}
 -\frac{b^2}{r^2} \frac{(2-\beta^2\sigma (r))(1+\sigma (r))}{2+\sigma (r)}
\eeq
A short calculation shows that $v^{(-)}$
is a monotonically decreasing function of $\beta$. Consequently,
we shall choose a convenient value
of it, for the sake of illustrating the features of the effective potential.
Figures (\ref{effpot1}) and (\ref{effpot2}) show the plots
of the potential for the $(-)$ metric
for several values of the relevant parameters.
\begin{figure}[h]
\begin{center}
\includegraphics[angle=-90,width=0.5\textwidth]{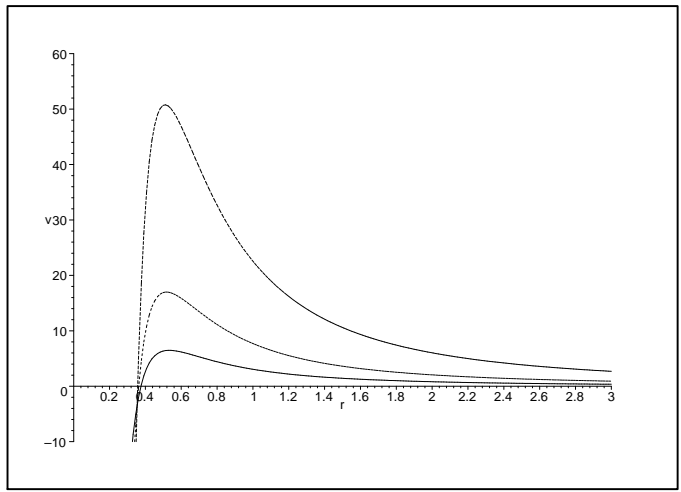}
\caption{Plot of the effective potential $v^{(-)}(r)$ for $q=1$,
$b=1,3,5$ (starting from the lowest curve), and $\beta=0.5$.}
\label{effpot1}
\end{center}
\end{figure}
\begin{figure}[h]
\begin{center}
\includegraphics[angle=-90,width=0.5\textwidth]{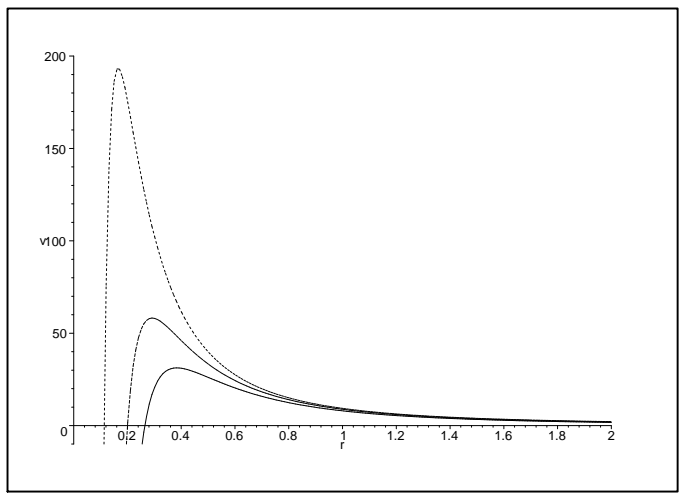}
\caption{Plot of the effective potential $v^{(-)} (r)$
for $b=3$, and $q=1, 3, 5$ (starting from the lowest
curve), and $\beta = 0.5$.}
\label{effpot2}
\end{center}
\end{figure}

The effective potential for the Gordon-like metric
can be obtained in the same way. From Eqns.(\ref{effpot}) and (\ref{gm}) we get
\beq
v^{(+)}(r) = 1 - \frac{2+\sigma (r)}{2}+ \frac{b^2}{2r^2}\;[2-\beta^2\sigma (r)].
\label{geffpot}
\eeq
The plots in Figures (\ref{effpot3}) and (\ref{effpot4})
show the dependence of $v^{(+)}(r)$ on the different parameters.
\begin{figure}[h]
\includegraphics[angle=-90,width=0.5\textwidth]{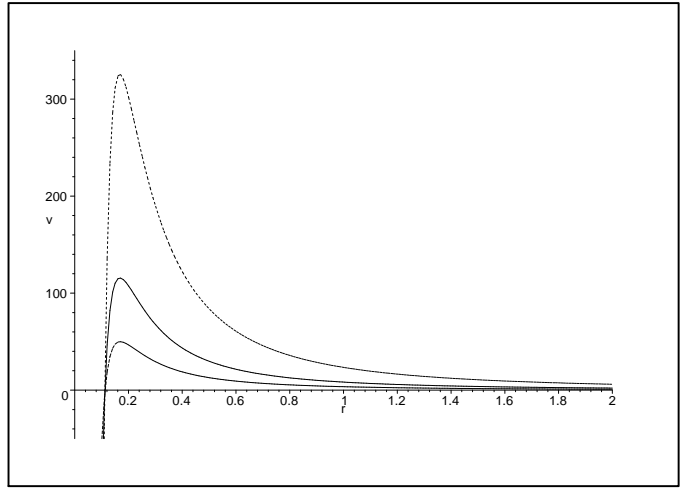}
\caption{Plot of the effective potential for the Gordon-like metric,
for $q=1$, $b=1, 3, 5$ (starting from the lowest curve), and $\beta = 0.5$.}
\label{effpot3}
\end{figure}

\begin{figure}[h]
\includegraphics[angle=-90,width=0.5\textwidth]{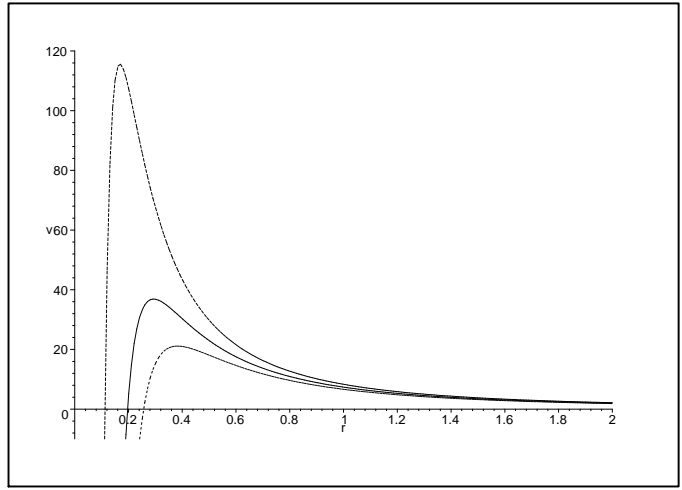}
\caption{Plot of the effective potential for the Gordon-like metric,
for $b=3$, $q=1, 3, 5$.}
\label{effpot4}
\end{figure}

We see from these plots that, in the case of a constant flux velocity,
the shape of the effective potential for both metrics qualitatively agrees with
that for photons moving on the geometry of a \Sch black hole
(see for instance Ref.\cite{wald}, pag. 143).

\section{Surface gravity and temperature}
\label{new}

Let us now go back to the more general case of $\beta = \beta(r)$,
and calculate the ``surface gravity'' of our analog black hole.
We present first the results for the constant permittivity case.
By setting
$\epsilon'(E)\equiv 0$ in the metrics Eqns.(\ref{h-2}) and (\ref{glmetric}),
we regain the example of constant
index of refraction studied for instance
in \cite{dbh}. It is easy to show from Eqn.(\ref{rh}) that the horizon of the black hole
in this case is given by
\beq
\beta^2(r_h) = \frac{1}{\bar\chi -1}.
\label{rad}
\eeq
The ``surface
gravity'' of
a spherically symmetric analog black hole in Schwarszchild coordinates
is given by \cite{viss}
\beq
\kappa = \frac {c^2}{ 2} \lim_{r\rightarrow r_h}\; \frac{g_{00,r}}{\sqrt{|g_{11}|
\;g_{00}}}.
\label{kappa}
\eeq
For the metrics Eqns.(\ref{h-2}) and (\ref{glmetric}) with
$\epsilon = \epsilon_0\bar\chi$ and $r_h$ given by
Eqn.(\ref{rad}), the analog surface gravity is
\beq
\kappa = - \frac{c^2}{2}\frac{1-\bar\chi}{\sqrt{\bar\chi}}\left.(\beta^2)'\right|_{r_h}.
\label{kappa1}
\eeq
This equation can be rewritten in terms of the velocity of light in the medium and the
refraction index, respectively given by
\beq
c_m^2 = \frac{1}{\mu_0\epsilon},\;\;\;\;\;\;\;\;\;\;\;\;\;\;\;
n= \frac{c}{c_m}.
\label{cm}
\eeq
The result is
\beq
\kappa = \frac{c^2}{2}\; \frac{1-n^2}{n}\; (\beta^2)_{,r}
\label{kappac}
\eeq
In this expression we can see the influence of the dielectric properties of the fluid
(through the index of refraction of the medium)
and also of its dynamics through the physical acceleration in the radial direction,
given by
$$
\left. a_r\right|_{r_h} = \frac{c^2}{2}\left.(\beta^2)'\right|_{r_h},
$$
for $\beta^2(r_h)\ll 1$. This acceleration
is a quantity that must
be determined solving the equations of motion of the fluid.
(Notice that if $\beta$ is set equal to 0 in Eqns.(\ref{g00})-(\ref{g11}), we cease to
have a black hole (this situation was analyzed in \cite{n1})).

Going back the the more general case of a linear permittivity, described
by the metrics given by Eqns.(\ref{nsm}) and (\ref{gm}), and considering that
$\beta(r_h)\ll 1$,
the
radius of the horizon is \footnote{Notice that we cannot take the limit $q\rightarrow
0$ in this expression or in any expression in which this one has been used.}:
\beq
r_h^2 = \frac{q\bar\chi^2}{4} \beta^4 (r_h).
\label{radius}
\eeq
Using the expressions given above,
the result for the surface gravity of the ``$-$'' black hole for $\beta(r_h)\ll 1$ is
\beq
\kappa^{(-)} = \left. \frac{c^2}{\beta}\left( \frac{1}{\bar\chi \sqrt q}  -
\frac 1 2 (\beta^2)'
\right)\right|_{r_h}.
\label{kappa2}
\eeq
This equation
differs from the surface gravity of the case of constant permittivity
(Eqn.(\ref{kappa1}))
by the presence of a new term that does not depend on the acceleration of the fluid.
To see where this new term comes from,
we can go back to the definition of the surface gravity given in Eqn.(\ref{kappa}),
and use the fact that in the high frequency limit the velocity of light and the index of
refraction in a medium of variable
$\epsilon$ are still given by Eqn.(\ref{cm}), replacing
the constant permittivity by $\epsilon = \epsilon (E)$. The result is
\beq
\kappa =\left.\left( \frac{c^2}{2}\;\frac{1-n^2(E)}{n(E)}\;(\beta^2)_{,r} +\frac{n(E)\epsilon(E)}
{\epsilon(E)+\epsilon(E)'E} \;(c^2_{m})_{,r}\right)\right|_{r_h}
\label{kappa3}
\eeq
In this expression, the first term is the generalization of the case $\epsilon=$ const.
(compare with Eqn.(\ref{kappac})), which mixes the acceleration of the fluid with
its dielectric properties. On the other hand, the second term, which is the
new term
displayed in Eqn.(\ref{kappa2}),
is related to the radial variation of the velocity of light in the medium.
It is important to point out that the result exhibited in Eqn.(\ref{kappa3})
is parallel
to
that of dumb holes: Unruh found in that case \cite{unruh} that the surface gravity
for constant speed of sound
is proportional to the acceleration of the fluid (as in the first term of
Eqn.(\ref{kappa3})).
This was generalized by Visser \cite{viss}, who showed that for a position-dependent
velocity of sound a second term appears, coming from the
gradients of the speed of sound, in analogy with the second term of Eqn.(\ref{kappa3}).

It is easy to show that the these results also apply to
the black hole described by the Gordon-like
metric. This is not surprising though, because of the conformal relation
between the two metrics, given by Eqn.(\ref{confinv}) \cite{jac}.\\[.1cm]

Let us remark once more that the concept of temperature, and indeed that of effective geometry
is valid in this context only for low-energy photons, {\em i.e.}
photons with wavelengths long compared to the
intermolecular spacing in the fluid.
For shorter wavelengths, there would be corrections to the propagation dictated by
the effective metric. However,
results
for other systems (such as dumb black holes \cite{unruh}
and Bose-Einstein condensates)
suggest that the phenomenon of Hawking
radiation is robust ({\em i.e.} independent of this "high-energy" physics). Consequently,
it makes sense to talk about the temperature of the radiation in these systems.

At first sight it may seem that by choosing an appropriate material and a convenient
value of the charge we could obtain a high value of the temperature of the radiation, given by
\beq
T\equiv \frac{\hbar }{2\pi k_B c}\;\kappa \approx 4\times 10^{-21}\;\kappa\; {\rm Ks}^2{\rm
/m}.
\label{temp}
\eeq

However, the equation for the surface gravity can be rewritten as
\footnote{Note that this equation depends on $\chi^{(2)}$ through the expression
for $r_h$, Eqn.(\ref{radius}).}
$$
\kappa = c^2 \left.\left(\frac{\beta}{2r} - \beta_{,r}
\right)\right|_{r_h}.
$$
We see then that, because $\beta(r_h)\ll 1$,
the new term appearing in $\kappa$ is bound to be very small. In spite of this result,
the emergence in the surface gravity of the term due to the variable velocity of light
 suggests that it may be worth to study if
some media with nonlinear dependence on an external electromagnetic field
can be used to generate analog black holes whose Hawking radiation could be
measured in laboratory.

\section*{Acknowledgements}

The authors would like to thank CNPq and FAPERj for financial support.

\end{document}